\documentclass[aps,groupedaddress]{revtex4}
\usepackage{epsfig}
\usepackage{graphicx}
\usepackage[T1]{fontenc}
\usepackage{ae}
\usepackage{color}
\usepackage[latin1]{inputenc}
\usepackage{amssymb,amsbsy,amsmath}
\usepackage{bbm}
\newtheorem{defi}{Definition}
\newtheorem{prop}{Proposition}

\newtheorem{theorem}{Theorem}

\begin{document}

%\begin{center}\today\end{center}

%%%%%%%%%%%%%%%%%%%%%%%%%%%%%%%%%%%%%%%%%%%%%%%%%%%%%%%%%%%%%%%%%%%%%%%%%%%%%

\title[DSM]
      {Dilation of stochastic matrices by coarse graining}
\author{Heinz-J\"urgen Schmidt$^1$
}
\address{$^1$  Universit\"at Osnabr\"uck,
Fachbereich Physik,
 D - 49069 Osnabr\"uck, Germany}

%\tableofcontents

\begin{abstract}
We consider two different ways of representing stochastic matrices by bi-stochastic ones acting on a larger probability space,
referred to as ``dilation by uniform coarse graining" and ``environmental dilation".
The latter is motivated by analogy to the dilation of operations in quantum theory.
Both types of dilation can be viewed as special cases of a general ``dilation by coarse graining".
We also discuss the entropy balance and illustrate our results, among others, by an example of a stochastic $4\times 4$-matrix,
which serves as a simplified model of the conditional action of Maxwell's demon.
\end{abstract}

\maketitle
%%%%%%%%%%%%%%%%%%%%%%%%%%%%%%%%%%%%%%%%%%%%%%%%%%%%%%%%%%%%%%%%%%%%%%%%%%%%%%%%%%%%%%%%%%%%%%%%%%%%%%%%%%%%%%%%%%%%%%%%%%%%%%%
\section{Introduction}\label{sec:Intro}
%%%%%%%%%%%%%%%%%%%%%%%%%%%%%%%%%%%%%%%%%%%%%%%%%%%%%%%%%%%%%%%%%%%%%%%%%%%%%%%%%%%%%%%%%%%%%%%%%%%%%%%%%%%%%%%%%%%%%%%%%%%%%%%
Stochastic matrices can be used to describe the conditional probability of transitions between different states of a system.
Because of the generality of this concept, there are numerous applications in various disciplines, from meteorology \cite{MW18}
and biology \cite{DK12} to physics \cite{CCMP05,SSG21}
and network theory, including Google's PageRank algorithm \cite{LM11}.

If the multiple applications of a stochastic matrix $T$ to a probability distribution converges then the limit must be
a fixed point of $T$, that is, an asymptotically stationary distribution.
If the uniform distribution is a fixed point of $T$, it is called a ``bi-stochastic matrix".
The convex set of bi-stochastic $N\times N$-matrices, the ``Birkhoff polytope", is not yet fully understood,
see, e.~g., \cite{P00,BP03,BE05}. Also the structural relations between stochastic matrices and bi-stochastic ones need to be further
analyzed. A first result of this kind is Sinkhorn's theorem \cite{S64,SK67}, saying that for every quadratic matrix $M$
with strictly positive entries there exist diagonal matrices $D_1, D_2$ with strictly positive elements such that
$S=D_1\,M\,D_2$ will be bi-stochastic.
The analogous result $T= M\,D_2$ for a stochastic matrices $T$ is trivial and hence in Sinkhorn's theorem
$M$ can be chosen as a stochastic matrix without loss of generality. The positivity condition of Sinkhorn's theorem
can be weakened to a certain degree \cite{BPS66, MO68}.

The present paper is motivated by physical applications. A bi-stochastic transition matrix $S$
in non-equilibrium statistical mechanics is the simplest case where a so-called ``Jarzinski equation" can be derived,
see  \cite{J97,K00}. More general cases where the Jarzynski equation holds can be obtained by a kind of coarse
graining of the set of elementary events, which leads to ``modified bi-stochastic matrices" \cite{SG20a,SG20b}.
We will call this process a ``dilation by uniform coarse graining", see Section \ref{sec:CG} for details.
``Coarse graining" is a general method of representing a physical system with fewer degrees of freedom than those
actually present in the system, see, e.~g, \cite{E04}.

Another physical context is given by the observation that stochastic matrices are, in some sense, special cases
of so-called ``operations" that occur in quantum measurement theory and quantum information theory, see \cite{K83,NC00,BLPY16}.
Operations are state changes that can be obtained by coupling the object system to some auxiliary system (environment)
and performing a time evolution in the total system followed by a reduction (partial trace) to the object system. This construction
leads to the mathematical notion of an ``environmental dilation" of the given operation. In the analogous environmental
dilation of a stochastic matrix the unitary time evolution will be replaced by a bi-stochastic time evolution of the total system
and the partial trace by the marginal probability distribution. A similar approach has been pursued by \cite{G10}
where a universal dilation of discrete Markov processes is constructed, however, by using an infinite state space for the
environment.

It is the purpose of this paper to make more precise the mentioned notions of ``dilation by uniform coarse graining"
and ``environmental dilation" of stochastic matrices and to show that they are special cases of a more general concept
of ``dilation by coarse graining". To this end we recapitulate in Section \ref{sec:GD} the general definitions concerning
stochastic matrices and present in Section \ref{sec:CG} the special definitions concerning uniform coarse graining.
According to Theorem \ref{T1} in Section \ref{sec:CG} every stochastic matrix with a rational fixed point admits a
dilation by uniform coarse graining. In order to extend this theorem to the general case we turn, in Section \ref{sec:DO},
to the dilation of quantum operations. After recalling the pertinent definitions in Section  \ref{sec:DOGD} we address
in Section \ref{sec:SMO} the above-mentioned relation between stochastic matrices $T$ and special operations of the form
$B=L\,A\,L$, $L$ being a so-called ``L\"uders operation". This relation leads to the ``environmental dilation" of $T$
considered in Section \ref{sec:DO1} and, as a special realization, its ``standard dilation" where the environment is modelled by
a copy of the object system. Section \ref{sec:C2D} is devoted to the proof that both kinds of dilation, that by uniform
coarse graining and the environmental dilation, are special cases of a general ``dilation by coarse graining".

The detailed example presented in Section \ref{sec:EX} is connected to the time-honored debate on Maxwell's demon
and similar interventions of ``intelligent beings" that may decrease the entropy of the system and thus seem to
violate the $2^{nd}$ law of thermodynamics, see \cite{EN98, EN99,LR03}. If the action of the demon can be descried by a stochastic matrix then its
environmental dilation resolves the apparent paradox. We close with a Summary in Section \ref{sec:SU}.

%%%%%%%%%%%%%%%%%%%%%%%%%%%%%%%%%%%%%%%%%%%%%%%%%%%%%%%%%%%%%%%%%%%%%%%%%%%%%%%%%%%%%%%%%%%%%%%%%%%%%%%%%%%%%%%%%%%%%%%%%%%%%%%%%%%%%%%%%%
\section{Dilation of stochastic matrices by uniform coarse graining}\label{sec:MR}
%%%%%%%%%%%%%%%%%%%%%%%%%%%%%%%%%%%%%%%%%%%%%%%%%%%%%%%%%%%%%%%%%%%%%%%%%%%%%%%%%%%%%%%%%%%%%%%%%%%%%%%%%%%%%%%%%%%%%%%%%%%%%%%%%%%%%%%%%%

%%%%%%%%%%%%%%%%%%%%%%%%%%%%%%%%%%%%%%%%%%%%%%%%%%%%%%%%%%%%%%%%%%%%%%%%%%%%%%%%%%%%%%%%%%%%%%%%%%%%%%%%%%%%%%%%%%%%%%%%%%%%%%%%%%%%%%%%%%
\subsection{General definitions}\label{sec:GD}
%%%%%%%%%%%%%%%%%%%%%%%%%%%%%%%%%%%%%%%%%%%%%%%%%%%%%%%%%%%%%%%%%%%%%%%%%%%%%%%%%%%%%%%%%%%%%%%%%%%%%%%%%%%%%%%%%%%%%%%%%%%%%%%%%%%%%%%%%%
Let ${\mathcal N}$ and  ${\mathcal M}$ be finite sets of size $N$ and $M$, resp.~, and
$T$ be an $N\times M$-matrix with non-negative entries. $T$ is called \textit{left-stochastic} iff
\begin{equation}\label{defleftstoch}
\sum_{n\in{\mathcal N}} T_{nm}=1\quad \mbox{for all } m\in{\mathcal M}
\;,
\end{equation}
\textit{right-stochastic} iff its transpose $T^\top$ is left-stochastic and
\textit{bi-stochastic} iff $T$ is both, left-stochastic and right-stochastic.
In the case of $T$ being bi-stochastic, it will necessarily be a square matrix, i.~e., $N=M$,
as can be easily seen by summation over all entries.
In accordance with the usage in the literature, by a ``stochastic matrix'',
for instance in the title of this paper, we always mean a left-stochastic square matrix,
since this is the most common case.

The convex set of probability distributions ${\mathbf p}$ can be identified with the $N-$simplex
\begin{equation}\label{defDelta}
  \Delta(N):=\left\{\mathbf{p}\in{\mathbbm R}^N\left| p_n\ge 0 \mbox{ for all } n\in{\mathcal N}\mbox{ and }\sum_{n\in{\mathcal N}}p_n=1\right.\right\}
  \;.
\end{equation}
Its open interior will be denoted by
\begin{equation}\label{defintDelta}
  \stackrel{\circ}{\Delta}(N):=\left\{\mathbf{p}\in\Delta(N)\left| p_n > 0 \mbox{ for all } n\in{\mathcal N}\right.\right\}
  \;.
\end{equation}

A stochastic matrix $T$ can be viewed as a linear map $T:{\mathbbm R}^N \rightarrow {\mathbbm R}^N$ that leaves the
affine hyperplane
\begin{equation}\label{defhyper}
 {\sf H}(N):=\left\{\mathbf{p}\in{\mathbbm R}^N\left|\sum_{n\in{\mathcal N}}p_n=1\right.\right\}
\end{equation}
and its subset $\Delta(N)$ invariant. Hence $T$ can also be viewed as an affine map
$T:\Delta(N) \rightarrow \Delta(N)$, and, conversely, any such map is given by a stochastic matrix.
We will make unrestricted use of this mathematical ambiguity if no misunderstandings are to be expected.

Every right-stochastic square matrix $T^\top$ has, by definition, the eigenvector
\begin{equation}\label{defe}
\mathbf{e}=(1,1,\ldots,1)^\top \in {\mathbbm R}^N
\end{equation}
corresponding to the eigenvalue $1$.
The corresponding left-stochastic matrix $T$ has the same eigenvalues as $T^\top$ and hence also a fixed point $\mathbf{p}$, i.~e., satisfying
$T\,\mathbf{p}=\mathbf{p}$.
If $T$ is \textit{irreducible}, then, by the theorem of Frobenius-Perron, $\mathbf{p}$ is unique (up to a factor), see, e.~g., \cite[Chapter III]{G05}.
In this case the fixed point will only have positive entries, i.~e., $\mathbf{p}\in  \stackrel{\circ}{\Delta}(N)$.

A bi-stochastic $N\times N$-matrix $S$ has the fixed point $\mathbf{p}=\frac{1}{N}\mathbf{e}\in  \stackrel{\circ}{\Delta}(N)$.
Hence, geometrically, $S$ does not only leave the subspace given by ${\mathbf e}\cdot {\mathbf p}=0$ invariant, but also its
one-dimensional orthogonal complement ${\mathbbm R}\, {\mathbf e}$. W.~r.~t.~an orthonormal basis adapted to these subspaces,
e.~g., the Fourier basis,
$S$ would assume the following block form
\begin{equation}\label{SONB}
  S\widehat{=}\left(
\begin{array}{cc}
 1 & {\mathbf 0}^\top \\
{\mathbf 0}& A \\
\end{array}
\right)
\;,
\end{equation}
whereas an $N\times N$-matrix $T$ that is only left stochastic would be transformed to
\begin{equation}\label{TONB}
  T\widehat{=}\left(
\begin{array}{cc}
 1 & {\mathbf 0}^\top \\
{\mathbf a}& A \\
\end{array}
\right)
\;.
\end{equation}

%%%%%%%%%%%%%%%%%%%%%%%%%%%%%%%%%%%%%%%%%%%%%%%%%%%%%%%%%%%%%%%%%%%%%%%%%%%%%%%%%%%%%%%%%%%%%%%%%%%%%%%%%%%%%%%%%%%%%%%%%%%%%%%%%%%%%%%%%%
\subsection{Uniform coarse graining}\label{sec:CG}
%%%%%%%%%%%%%%%%%%%%%%%%%%%%%%%%%%%%%%%%%%%%%%%%%%%%%%%%%%%%%%%%%%%%%%%%%%%%%%%%%%%%%%%%%%%%%%%%%%%%%%%%%%%%%%%%%%%%%%%%%%%%%%%%%%%%%%%%%%

In this paper we will consider a somewhat reduced concept of ``coarse graining'', compared with \cite{E04},
which consists simply of a partition of a finite set. Hence let ${\mathcal D}$ be a finite set of size $d\ge N$ and consider the
partition
\begin{equation}\label{defpart}
 {\mathcal D}= \biguplus_{n\in{\mathcal N}}{\mathcal D}_n
\end{equation}
of ${\mathcal D}$ into disjoint subsets (or ``equivalence classes") ${\mathcal D}_n$.
If at least one subset ${\mathcal D}_n$ contains more than one element, we will speak of a ``proper coarse graining".
Let $d_n, \, n\in{\mathcal N},$  denote the number of elements of ${\mathcal D}_n$ and hence
\begin{equation}\label{sumdn}
 \sum_{n\in{\mathcal N}} d_n = \sum_{n\in{\mathcal N}} \left|{\mathcal D}_n \right| = \left|{\mathcal D}\right|=d
 \;.
\end{equation}

The characteristic functions of the ${\mathcal D}_n,\, n\in {\mathcal N},$ are
\begin{equation}\label{defcharf}
  \chi^{(n)}_\nu :=\left\{
  \begin{array}{r@{\quad : \quad}l}
  1& \nu\in {\mathcal D}_n ,\\
    0& \mbox{else},
  \end{array}
  \right.
\end{equation}
for all $\nu\in{\mathcal D}$, satisfying
\begin{equation}\label{charf1}
  \sum_{n\in{\mathcal N}} \chi^{(n)}_\nu = 1 \quad \mbox{for all } \nu\in{\mathcal D}
  \;,
\end{equation}
and
\begin{equation}\label{charf2}
  \sum_{\nu\in{\mathcal D}} \chi^{(n)}_\nu = d_n \quad \mbox{for all } n\in{\mathcal N}
  \;.
\end{equation}

\begin{figure}[t]
\centering
\includegraphics[width=0.7\linewidth]{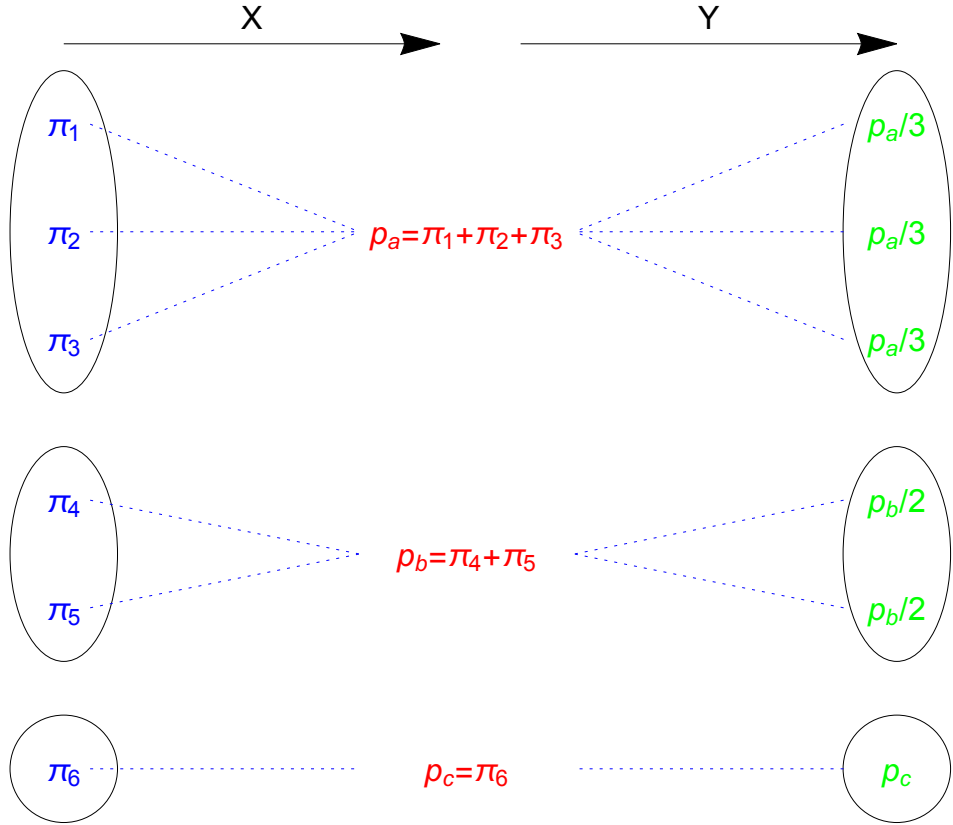}
\caption{Visualization of the map $X$ describing coarse graining and its uniform right inverse $Y$.
In this example, ${\mathcal D}=\{1,2,\ldots,6\}$, ${\mathcal N}=\{a,b,c\}$ and the partition
${\mathcal D}={\mathcal D}_a \cup {\mathcal D}_b \cup {\mathcal D}_c  $ is indicated by three ellipses.
}
\label{FIGGR}
\end{figure}

If ${\boldsymbol \pi}\in \Delta (d)$ is viewed as a probability distribution over the set of elementary events ${\mathcal D}$ then
it is possible to perform a partial summation of the probabilities  over the subsets ${\mathcal D}_n$ thereby obtaining
a ``coarse grained" probability distribution $\mathbf{p}\in \Delta (N)$, see Figure \ref{FIGGR}.
This process can be represented by a surjective affine map
\begin{equation}\label{mapX}
  X:\Delta(d) \rightarrow \Delta(N)
  \;,
\end{equation}
with (left stochastic) matrix representation
\begin{equation}\label{matX}
  X_{n\nu}:= \chi^{(n)}_\nu,\quad \mbox{for all } n\in{\mathcal N} \mbox{ and }\nu\in{\mathcal D}
  \;,
\end{equation}
such that
\begin{equation}\label{Xpi}
  p_n = \left( X\,{\boldsymbol \pi}\right)_n = \sum_\nu X_{n\nu}\,\pi_\nu \stackrel{(\ref{matX})}{=} \sum_\nu \chi^{(n)}_\nu\,\pi_\nu
  \stackrel{(\ref{defcharf})}{=}
  \sum_{\nu\in{\mathcal D}_n}\pi_\nu
  \;,
\end{equation}
in accordance with the above prescription of coarse graining probability distributions.

Proper coarse graining decreases the Shannon entropy of a probability distribution:
\begin{prop}\label{PS}
\begin{equation}\label{entdec}
  H({\mathbf p})=H(X\,{\boldsymbol\pi})< H({\boldsymbol\pi})
\end{equation}
for all ${\boldsymbol\pi}\in\Delta(d)$, where the Shannon entropy is defined as
\begin{equation}\label{defent}
H({\mathbf p}):=-\sum_n
\left\{
  \begin{array}{r@{\quad : \quad}l}
 p_n\,\log p_n &p_n>0,\\
  0& p_n=0,
  \end{array}
  \right.
\;,
\end{equation}
following \cite{S48} up to the choice of units,
and the coarse graining $ {\mathcal D}= \biguplus_{n\in{\mathcal N}}{\mathcal D}_n $ is assumed to be proper.
\end{prop}
{\bf Proof}:
For all $n\in{\mathcal N}$ we have
\begin{eqnarray}
\label{dec1}
 p_n\,\log p_n&\stackrel{(\ref{Xpi})}{=}&\left( \sum_{\nu\in{{\mathcal D}_n}}\pi_\nu\right)\,\log\left( \sum_{\mu\in{{\mathcal D}_n}}\pi_\mu\right) \\
 \label{dec2}
   &\ge&  \sum_{\nu\in{{\mathcal D}_n}}\pi_\nu\,\log \pi_\nu
   \;,
\end{eqnarray}
using that $\log$ is a strictly monotonic function. Since the coarse graining is assumed to be proper we may replace the $\ge$ in (\ref{dec2})
by $>$ for, at least, one index $n\in{\mathcal N}$. From this follows $H({\mathbf p})< H({\boldsymbol\pi})$,
where the minus sign in (\ref{defent}) is to be noted.              \hfill$\Box$\\

In order to define coarse grained version of a stochastic $d\times d-$ matrix $S$ it is not sufficient
to know how to ``project" the probability distributions according to the affine map $X:\Delta(d)\rightarrow\Delta(N)$
but we also need the inverse process of ``lifting" a probability distribution from $\Delta(N)$ to $\Delta(d)$.
We hence need to define an affine map
$Y:\Delta(N) \rightarrow \Delta(d)$ that is a right-inverse (or ``section")
of $X$, i.~e., satisfying $X\,Y=\mathbbm{1}$.
In general, $X$ has many right inverses; in this Section we select one of these by choosing the matrix representation
\begin{equation}\label{defY}
 Y_{\nu n}:= \frac{1}{d_n}\chi^{(n)}_\nu,\quad \mbox{for all } \nu\in{\mathcal D} \mbox{ and }n\in{\mathcal N}
  \;.
\end{equation}
Indeed, this implies
\begin{equation}\label{XY}
\left( X\,Y\right)_{nm}=\sum_\nu X_{n\nu}\,Y_{\nu m} \stackrel{(\ref{matX},\ref{defY})}{=}
\sum_\nu \chi^{(n)}_\nu\,\frac{1}{d_n}\,\chi^{(m)}_\nu\stackrel{(\ref{defcharf})}{=}\delta_{nm}
\;,
\end{equation}
and hence $Y$ is a right inverse of $X$, see Figure \ref{FIGGR}.
It will be called the ``uniform right inverse" of $X$ in what follows.

The above definition (\ref{defY}) of the uniform right inverse $Y$
means that we choose the unknown probabilities $\pi_\nu,\,\nu\in {\mathcal D}_n$ with given sum $p_n$
as the uniform mean value $\pi_\nu= \frac{p_n}{d_n}$. This follows from
\begin{equation}\label{Yp}
 \left( Y\,\mathbf{p}\right)_\nu = \sum_m Y_{\nu m} p_m \stackrel{(\ref{defY})}{=} \sum_m \frac{1}{d_m}\,\chi^{(m)}_\nu \, p_m
 \stackrel{(\ref{defcharf})}{=}\frac{p_n}{d_n}
 \quad \mbox{if }\nu \in {\mathcal D}_n
 \;.
\end{equation}
The product $Y\,X:\Delta(d) \rightarrow \Delta(d)$ replaces the probability distribution ${\boldsymbol\pi}$ by its
averaged distribution over the subsets ${\mathcal D}_n$, see Figure \ref{FIGGR}. It is idempotent since
$\left(Y\,X\right)^2= Y\,\left( X\,Y\right)\,X =Y\,{\mathbbm 1}\,X=Y\,X$.

Let $S$ be a $d\times d-$stochastic matrix, $S:\Delta(d)\rightarrow\Delta(d)$,
and define its coarse grained version by
\begin{equation}\label{defT}
  T:=X\,S\,Y
  \;,
\end{equation}
in components,
\begin{equation}\label{defTcom}
  T_{nm}=\sum_{\nu\mu}X_{n\nu}\,S_{\nu\mu}\,Y_{\mu m}\stackrel{(\ref{matX},\ref{defY})}{=}
  \sum_{\nu\mu}\chi^{(n)}_\nu\,\chi^{(m)}_\mu\,\frac{1}{d_m}\,S_{\nu\mu}
  \;.
\end{equation}
Clearly, $T_{nm}\ge 0$ for all $n,m\in{\mathcal N}$. Moreover,
\begin{eqnarray}\label{Tstoch1}
  \sum_n T_{nm}&\stackrel{(\ref{defT})}{=}& \sum_{\nu\mu}\underbrace{\left(\sum_n \chi^{(n)}_\nu\right)}_{\stackrel{(\ref{charf1})}{=}1}\,
  \chi^{(m)}_\mu\,\frac{1}{d_m}\,S_{\nu\mu}\\
  \label{Tstoch2}
  &=& \underbrace{\left(\sum_\mu \chi^{(m)}_\mu\frac{1}{d_m}\right)}_{\stackrel{(\ref{charf2})}{=}1}\,
   \underbrace{\left(\sum_{\nu} S_{\nu\mu}\right)}_{=1}=1
     \;.
\end{eqnarray}
This shows that $T$ is indeed a stochastic $N\times N-$matrix and hence can be viewed as an affine map $T:\Delta(N)\rightarrow \Delta(N)$.

We emphasize that, in general, coarse graining is not compatible with the multiplication of stochastic matrices.
Hence the above construction is not automatically applicable to \textit{Markov chains}, see also \cite[Chapt. III]{B16}.

Now assume that $S$ has a fixed point $\mathbf{q}\in\Delta(d)$ that lies in the range of $Y$, i.~e.,
$S\,\mathbf{q}=\mathbf{q}$ and $\mathbf{q}=Y\,\mathbf{p}$ for some $\mathbf{p}\in \Delta(N)$.
Then it follows that
\begin{equation}\label{fixT}
 T\,\mathbf{p}= X\,S\,Y\,\mathbf{p}=X\,S\,\mathbf{q}=X\,\mathbf{q}=\underbrace{X\,Y}_{\mathbbm 1}\,\mathbf{p}=\mathbf{p}
 \;,
\end{equation}
i.~e., $\mathbf{p}$ will be a fixed point of $T$.

We will apply the latter result to the case where $S$ is bi-stochastic and has the fixed point $\mathbf{q}=\frac{1}{d}\mathbf{E}$,
where
\begin{equation}\label{defE}
  \mathbf{E}:=(1,1,\ldots,1)^\top \in \mathbbm{R}^d
  \;.
\end{equation}
In this case $\mathbf{q}=Y\,\mathbf{p}$ with $p_n:=\frac{d_n}{d}$ for all $n\in{\mathcal N}$. This follows from
\begin{equation}\label{Yp}
  \left(Y\,\mathbf{p}\right)_\nu = \sum_n Y_{\nu n}\,p_n \stackrel{(\ref{defY})}{=}\sum_n \frac{1}{d_n}\,\chi^{(n)}_\nu\,
  \frac{d_n}{d}= \frac{1}{d}\sum_n \chi^{(n)}_\nu\stackrel{(\ref{charf1})}{=}\frac{1}{d}=q_\nu
  \;.
\end{equation}
It follows that $\mathbf{p}$ is a fixed point of $T$ with rational components.

%%%%%%%%%%%%%%%%%%%%%%%%%%%%%%%%%%%%%%%%%%%%%%%%%%%%%%%%%%%%%%%%%%%%%%%%%%%%%%%%%%%%%%%%%%%%%%%%%%%%%%%%%%%%%%%%%%%%%%%%%%%%%%%%%%%%%%%%%%
\subsection{Dilation by uniform coarse graining}\label{sec:DCG}
%%%%%%%%%%%%%%%%%%%%%%%%%%%%%%%%%%%%%%%%%%%%%%%%%%%%%%%%%%%%%%%%%%%%%%%%%%%%%%%%%%%%%%%%%%%%%%%%%%%%%%%%%%%%%%%%%%%%%%%%%%%%%%%%%%%%%%%%%%
We have seen in the last Subsection \ref{sec:CG} that, starting with a bi-stochastic matrix $S$, by means of uniform coarse graining we obtain a stochastic matrix
$T$ possessing a fixed point $\mathbf{p}$  with rational components (or, shortly, a {\it rational fixed point}).
Here we consider the inverse problem: Can every stochastic matrix $T$ with a rational fixed point $\mathbf{p}$ be obtained in this way?
In this case we will speak of a ``dilation of $T$ by uniform coarse graining", which refers to the choice of the
\textit{uniform} right inverse $Y$ of $X$. The answer is given by the following:
\begin{theorem}\label{T1}
  Given a stochastic $N\times N-$matrix $T$ with a rational fixed point  $\mathbf{p}$,
  then there exists a bi-stochastic $d\times d$-matrix $S$ such that $T$ is obtained from $S$ by means of uniform coarse graining.
\end{theorem}
{\bf Proof}: Since the components of  $\mathbf{p}$ are rational numbers they can be written in the form
$p_n=\frac{d_n}{d}$ with $d_n\in{\mathbbm N},\,d_n>0$ for $n\in{\mathcal N}$ and $d=\sum_n d_n$ being the least common denominator of the $p_n$.
Let ${\mathcal D}=\{1,2,\ldots,d\}$ and ${\mathcal D}=\biguplus_{n\in{\mathcal N}}{\mathcal D}_n$ be a partition
of ${\mathcal D}$ satisfying $\left| {\mathcal D}_n\right|=d_n$ for all $n\in{\mathcal N}$. Choose $X:\Delta(d)\rightarrow\Delta(N)$
according to (\ref{matX}) and its uniform right inverse $Y$ according to (\ref{defY}). Define $S:\Delta(d)\rightarrow\Delta(d)$ by
$S:= Y\,T\,X$. If follows that
\begin{equation}\label{XSY}
 X\,S\,Y=X\,\left(Y\,T\,X\right)\,Y =\underbrace{\left( X\,Y\right)}_{\mathbbm 1}\,T\,\underbrace{\left( XY\right)}_{\mathbbm 1}=T
 \;.
\end{equation}
Hence $T$ is obtained from $S$ by means of uniform coarse graining and it remains to show that $S$ is bi-stochastic.

To this end note that the entries of $S$ are non-negative and consider
\begin{eqnarray}
\label{Snumu1}
  \sum_\nu S_{\nu\mu} &=& \sum_{\nu n m} Y_{\nu n}\,T_{nm}\,X_{m\mu}\\
  \label{Snumu2}
  &\stackrel{(\ref{matX},\ref{defY})}{=}& \underbrace{\sum_\nu\left(\frac{1}{d_n}\chi^{(n)}_\nu \right)}_{\stackrel{(\ref{charf2})}{=}1}
  \underbrace{\sum_m \chi^{(m)}_\mu}_{\stackrel{(\ref{charf1})}{=}1}
  \underbrace{\sum_n T_{nm}}_{=1}\\
  \label{Snumu3}
  &=&1
  \;,
\end{eqnarray}
as well as
\begin{eqnarray}
\label{Snumu4}
  \sum_\mu S_{\nu\mu} &=& \sum_{\mu n m} Y_{\nu n}\,T_{nm}\,X_{m\mu}\\
  \label{Snumu5}
  &\stackrel{(\ref{matX},\ref{defY})}{=}&\sum_n \frac{1}{d_n}\chi^{(n)}_\nu\,
  \sum_m T_{nm}\,
    \underbrace{\sum_\mu \chi^{(m)}_\mu}_{\stackrel{(\ref{charf2})}{=}d_m}
    \\
  \label{Snumu6}
  &=&\sum_n \frac{1}{{d}_n}\chi^{(n)}_\nu\,{d}_n\stackrel{(\ref{charf1})}{=}1
  \;,
\end{eqnarray}
where in (\ref{Snumu6}) we have used that  $\sum_m T_{nm}d_m=d_n$ since ${\mathbf p}$ with components $p_n=\frac{d_n}{d}$ is a fixed point of $T$.
\hfill$\Box$\\

The problem remains whether dilations of stochastic matrices are possible without the assumption of a rational fixed point.

%%%%%%%%%%%%%%%%%%%%%%%%%%%%%%%%%%%%%%%%%%%%%%%%%%%%%%%%%%%%%%%%%%%%%%%%%%%%%%%%%%%%%%%%%%%%%%%%%%%%%%%%%%%%%%%%%%%%%%%%%%%%%%%%%%%%%%%%%%
\section{Dilation of operations}\label{sec:DO}
%%%%%%%%%%%%%%%%%%%%%%%%%%%%%%%%%%%%%%%%%%%%%%%%%%%%%%%%%%%%%%%%%%%%%%%%%%%%%%%%%%%%%%%%%%%%%%%%%%%%%%%%%%%%%%%%%%%%%%%%%%%%%%%%%%%%%%%%%%
To address the problem  mentioned at the end of the last Section we will recall the dilation theory of operations in quantum theory
outlined, e.~g., in  \cite{K83,NC00,BLPY16} and apply this to stochastic matrices.

%%%%%%%%%%%%%%%%%%%%%%%%%%%%%%%%%%%%%%%%%%%%%%%%%%%%%%%%%%%%%%%%%%%%%%%%%%%%%%%%%%%%%%%%%%%%%%%%%%%%%%%%%%%%%%%%%%%%%%%%%%%%%%%%%%%%%%%%%%
\subsection{General Definitions}\label{sec:DOGD}
%%%%%%%%%%%%%%%%%%%%%%%%%%%%%%%%%%%%%%%%%%%%%%%%%%%%%%%%%%%%%%%%%%%%%%%%%%%%%%%%%%%%%%%%%%%%%%%%%%%%%%%%%%%%%%%%%%%%%%%%%%%%%%%%%%%%%%%%%%

Let ${\mathcal H}$ be a  finite-dimensional complex Hilbert space,
$B({\mathcal H})$ denote the space of Hermitean operators  $A:{\mathcal H}\longrightarrow {\mathcal H}$
and $B^+({\mathcal H})$ the cone of positively semi-definite operators, i.~e., having only non-negatives eigenvalues.
The convex subset $B_1^+({\mathcal H})\subset B^+({\mathcal H})$ consists of \textit{statistical operators}  $\rho$
with $\mbox{Tr} \rho=1$. Such operators physically describe (mixed) states. \textit{Pure states} are
represented by one-dimensional projectors $P_\psi$, where $\psi\in{\mathcal H}$ with $\|\psi\|=1$.

According to \cite[8.2.1]{NC00},  there are three equivalent ways to define \textit{operations}:
\begin{itemize}
  \item By considering the system coupled to environment,
  \item by an operator-sum representation, or
  \item via physically motivated axioms.
\end{itemize}
Here we follow the second approach and define an ``operation" to be a map
$A:B({\mathcal H})\longrightarrow B({\mathcal H})$ of the form
\begin{equation}\label{OI2}
 A( \rho)= \sum_{i\in{\mathcal I}}A_i\,\rho\,A_i^\ast
  \;,
\end{equation}
with the linear \textit{Kraus operators} $A_i:{\mathcal H}\rightarrow{\mathcal H}$ and a finite
index set ${\mathcal I}$,
such that
\begin{equation}\label{OI2a}
  \mbox{Tr }A( \rho) \le 1 \quad \mbox{for all } \rho\in B_1^+({\mathcal H})
  \;,
\end{equation}
see \cite{K83}. It follows that an operation is linear
and maps $B^+({\mathcal H})$ into itself.
It is mathematically convenient not to require that an operation preserves the trace.
The normalized state after the operation would be obtained as $\frac{A( \rho)}{\mbox{Tr }A( \rho)}$.
Obviously, any product (concatenation) of operations is again an operation.

Operations are intended to describe state changes due to measurements. For example, the total \textit{L\"uders operation}
\begin{equation}\label{LO}
  \rho\mapsto L(\rho):=\sum_{n\in{\mathcal N}}P_n\,\rho\,P_n
\end{equation}
where $\left( P_n\right)_{n\in{\mathcal N}}$ is a complete family of mutually orthogonal projections in ${\mathcal H}$
is a trace-preserving operation in the above sense with
${\mathcal I}={\mathcal N}$ and $A_n=P_n$ for all $n\in{\mathcal N}$.
It models a special state transformation after the measurement of an observable given by a self-adjoint operator
with eigenprojections $P_n, \,n\in{\mathcal N}$, see \cite{BLPY16}.

There exists a so-called \textit{statistical duality} between states and observables, see \cite{BLPY16}, chapter 23.1.
In the finite-dimensional case $B({\mathcal H})$ can be identified with its dual space $B({\mathcal H})^\ast$
by means of the Euclidean scalar product $\mbox{Tr}\,(A\,B)$. Physically, we may distinguish between the two
spaces in the sense that $B({\mathcal H})$ is spanned by the subset of statistical operators representing states
and $B({\mathcal H})^\ast$  is spanned by the subset of operators with eigenvalues in the interval $[0,1]$ representing
\textit{effects}. Effects describe yes-no-measurements including the subset of projectors, which are the extremal points of the
convex set of effects, see \cite{BLPY16}.

Every operation $A:B({\mathcal H})\rightarrow B({\mathcal H})$, viewed as a transformation of states (Schr\"odinger picture)
gives rise to the dual operation $A^\ast:B({\mathcal H})^\ast\longrightarrow B({\mathcal H})^\ast$
viewed as a transformation of effects (Heisenberg picture).
Reconsider the representation  (\ref{OI2}) of the operation $A$ by means of the Kraus operators $A_i$. Then the dual
operation $A^\ast$ has the corresponding representation
\begin{equation}\label{M1}
 A^\ast(X)= \sum_{i\in{\mathcal I}}A_i^\ast\,X\,A_i
  \;,
\end{equation}
for all $X\in B({\mathcal H})^\ast$.

The condition that the operation (\ref{OI2}) will
be trace-preserving translates into
\begin{equation}\label{M3}
\sum_{i\in{\mathcal I}_n} A_{i}^\ast\, A_{i}
 ={\mathbbm 1}
 \;.
\end{equation}
Thus every dual operation yields a resolution of the identity by means of effects
\begin{equation}\label{eff1}
F_i:= A_{i}^\ast\, A_{i},\quad i\in{\mathcal I}
\;,
\end{equation}
and hence to a generalized observable in the sense of a \textit{positive operator-valued measure}
$F=\left( F_n\right)_{n\in{\mathcal N}}$, see \cite{BLPY16}. Note, however, that compared to the general
definition in \cite{BLPY16} we will have to consider generalized observables only in the discrete, finite-dimensional case.
The traditional notion of ``sharp"  observables represented by self-adjoint operators corresponds to the special case
of a projection-valued measure $\left(P_n\right)_{n\in {\mathcal N}}$ satisfying $\sum_{n\in{\mathcal N}}P_n={\mathbbm 1}$.

%%%%%%%%%%%%%%%%%%%%%%%%%%%%%%%%%%%%%%%%%%%%%%%%%%%%%%%%%%%%%%%%%%%%%%%%%%%%%%%%%%%%%%%%%%%%%%%%%%%%%%%%%%%%%%%%%%%%%%%%%%%%%%%%%%%%%%%%%
\subsection{Stochastic matrices as special operations}\label{sec:SMO}
%%%%%%%%%%%%%%%%%%%%%%%%%%%%%%%%%%%%%%%%%%%%%%%%%%%%%%%%%%%%%%%%%%%%%%%%%%%%%%%%%%%%%%%%%%%%%%%%%%%%%%%%%%%%%%%%%%%%%%%%%%%%%%%%%%%%%%%%%%

We consider the product $B=L\,A\,L$ of the special L\"uders operation $L$ defined in (\ref{LO}) followed by a general operation
$A$ of the form (\ref{OI2}) and finally again by $L$. The Kraus operator representation of $B$ will hence be given by
\begin{equation}\label{KrausB}
  B(\rho)=\sum_{n,m\in{\mathcal N},i\in{\mathcal I}}P_m\,A_i\,P_n\,\rho\,P_n\,A_i^\ast\,P_m
  \;.
\end{equation}
We assume that the projections $P_n$ are one-dimensional, $P_n=|n\rangle \langle n|$, for all $n\in{\mathcal N}$,
using Dirac's bra-ket notation, and hence
$N=\left| {\mathcal N}\right|$ will be the dimension of the Hilbert space ${\mathcal H}$.
Moreover,
\begin{equation}\label{TrrhoPn}
  P_n\,\rho\,P_n=\mbox{Tr}\left( \rho\,P_n\right)\,P_n=:p_n\,P_n,\quad\mbox{such that } \sum_{n\in{\mathcal N}} p_n=\mbox{Tr } \rho=1
  \;.
\end{equation}
Then it follows that
\begin{eqnarray}
\label{Brho1}
   B(\rho) &\stackrel{(\ref{KrausB},\ref{TrrhoPn})}{=}&\sum_{n,m,i}P_m\,A_i\left(p_n \,P_n \right)A_i^\ast\,P_m \\
   \label{Brho2}
   &=&\sum_{n,m,i} p_n\,| m\rangle\langle m| A_i | n\rangle \langle n| A_i^\ast| m\rangle\langle m |\\
   \label{Brho3}
   &=& \sum_{n,m} T_{mn}\,p_n\,P_m
   \;,
\end{eqnarray}
where
\begin{equation}\label{defTmn}
 T_{mn}:= \sum_i \langle m| A_i | n\rangle \langle n| A_i^\ast| m\rangle=\sum_i \left| \left\langle m\left| A_i \right| n\right\rangle\right|^2
 \;.
\end{equation}
Obviously, $T_{mn}\ge 0$ for all $m,n\in{\mathcal N}$ and
\begin{eqnarray}
\label{Tstoch1}
 \sum_m T_{mn} &\stackrel{(\ref{defTmn})}{=}& \sum_{m,i} \langle m| A_i | n\rangle \langle n| A_i^\ast| m\rangle
 = \sum_i \mbox{Tr}\left( A_i \,P_n \,A_i^\ast\right)\\
 \label{Tstoch2}
   &=&\mbox{Tr}\left(\underbrace{\left(\sum_i A_i^\ast\,A_i\right)}_{\stackrel{(\ref{M3})}{=}{\mathbbm 1}}\,P_n\right)=\mbox{Tr}\,P_n=1
   \;,
\end{eqnarray}
for all $n\in{\mathcal N}$
and hence the matrix $T$ with entries (\ref{defTmn}) will be a stochastic $N\times N$-matrix.
In Eq.~(\ref{Tstoch2}) we have used that the trace of a product of operators is invariant under cyclic permutations.
Thus every operation of the form $B=L\,A\,L$ yields a stochastic matrix $T$ via (\ref{defTmn}).

Conversely, for every stochastic $N\times N$-matrix $T$ we may find an operation of the form $\widetilde{B}=L\,\widetilde{A}\,L$ that yields $T$ in the above sense by choosing
${\mathcal I}={\mathcal N}$ and
\begin{equation}\label{T2B}
 \widetilde{A}_i\,|n\rangle:= \sqrt{T_{in}} \,|i\rangle
\end{equation}
for all $n,i\in{\mathcal N}$.  This follows from
\begin{eqnarray}
\label{T2B1}
 \widetilde{T}_{mn} &\stackrel{(\ref{defTmn})}{=}& \sum_i \langle m |\widetilde{A}_i|\,n\rangle \langle n | \widetilde{A}_i^\ast| m\rangle\\
 \label{T2B2}
  &\stackrel{(\ref{T2B})}{=}& \sum_i \langle m |\sqrt{T_{in}}|i\rangle \langle  i \sqrt{T_{in}} | m\rangle\\
  \label{T2B3}
  &=& \sum_i T_{in}\, \underbrace{\langle m | i\rangle}_{\delta_{mi}}\langle i | m\rangle = T_{mn}
  \;.
\end{eqnarray}

The relation between stochastic matrices and operations has also been considered in \cite{Z14}.

%%%%%%%%%%%%%%%%%%%%%%%%%%%%%%%%%%%%%%%%%%%%%%%%%%%%%%%%%%%%%%%%%%%%%%%%%%%%%%%%%%%%%%%%%%%%%%%%%%%%%%%%%%%%%%%%%%%%%%%%%%%%%%%%%%%%%%%%%%
\subsection{Environmental dilation of operations}\label{sec:DO1}
%%%%%%%%%%%%%%%%%%%%%%%%%%%%%%%%%%%%%%%%%%%%%%%%%%%%%%%%%%%%%%%%%%%%%%%%%%%%%%%%%%%%%%%%%%%%%%%%%%%%%%%%%%%%%%%%%%%%%%%%%%%%%%%%%%%%%%%%%%

We recall the first approach to operations according to \cite[8.2.1]{NC00}, which is based on the coupling of the object system to some environment.
Let ${\mathcal K}$ the Hilbert space of the environment and $\rho_2$ some statistical operator in ${\mathcal K}$ describing the initial
state of the environment, such that the total initial state will be $\rho\otimes \rho_2$. After some unitary time evolution described by
the operator $U$ defined on ${\mathcal H}\otimes{\mathcal K}$ we consider the state $\rho'$ reduced to the object system and defined by the partial trace
\begin{equation}\label{rhoprime}
\rho'= \mbox{Tr}_2\left(U\left( \rho\otimes \rho_2\right) U^\ast\right)
\;.
\end{equation}
It can be shown that $\rho\mapsto \rho'$ is a trace-preserving operation and, vice versa, that every trace-preserving operation
can be written in the form of (\ref{rhoprime}).

The details for the latter claim are as follows. Assume a trace-preserving operation $A$ of the form (\ref{OI2}),
hence satisfying (\ref{M3}).
Then we may choose ${\mathcal K}={\mathbbm C}^{\mathcal I}$,
with $\left| i\right\rangle_{i\in{\mathcal I}}$ being an orthonormal base in ${\mathcal K}$, and $\rho_2$ as the pure state
$\rho_2=\left| 0\right\rangle\left\langle 0\right|$ with some $0\in{\mathcal I}$ which may be chosen, without loss of generality,
as the first element of ${\mathcal I}$ if ${\mathcal I}$ has natural order. The unitary operator $U$ will be defined as the
extension of the partial isometry $V: {\mathcal H}\otimes |0\rangle \rightarrow  {\mathcal H}\otimes {\mathcal K}$ defined by
\begin{equation}\label{defV}
  V \left| n,0\right\rangle:= V \left| n\right\rangle\otimes \left| 0 \right\rangle :=
  \sum_{i\in {\mathcal I}}A_i \left| n \right\rangle \otimes  \left| i \right\rangle
  \;.
\end{equation}
Then it can be shown that
\begin{equation}\label{Dila}
  A(\rho)= \mbox{Tr}_2\left(U\left( \rho\otimes |0\rangle \langle 0|\right) U^\ast\right)
  \;,
\end{equation}
for all $\rho\in{\mathcal B}_1^+({\mathcal H})$,
see see \cite[Eq.~8.3.9]{NC00} or \cite[Appendix A]{S21}.
We will refer to the special realization of a dilation sketched in this paragraph as the ``standard dilation" of the operation $A$.\\

The dilation concept for operations can be utilized for the analogous problem of dilation of stochastic matrices in two ways:
Firstly, we may define an analogous concept of dilation for stochastic matrices.
\begin{defi}\label{D3}
Let $T$ be a stochastic $N\times N$-matrix. Then an ``environment dilation" of $T$ consists of a triple $\left({\mathcal M}, {\boldsymbol\rho}, R \right)$,
where ${\mathcal M}$ is a finite set of size $M$, ${\boldsymbol \rho}\in\Delta(M)$ a probability distribution and
$R:\Delta(N\times M)\rightarrow\Delta(N\times M)$
a bi-stochastic matrix, such that
\begin{equation}\label{D3dil}
  T {\mathbf p}= {\sf M}_1\left( R\, {\mathbf p}\otimes {\boldsymbol\rho}\right)
\end{equation}
for all ${\mathbf p}\in \Delta(N)$. Here ${\mathbf p}\otimes{\boldsymbol\rho}$ denotes the initial product probability distribution
and ${\sf M}_1(\ldots)$ the first marginal distribution.
\end{defi}
The coordinate version of (\ref{D3dil}) reads:
\begin{equation}\label{D3coor}
 \sum_n T_{mn}\,p_n =  \sum_{ij,n}   R_{mi,nj}\,p_n\,\rho_j
 \;,
\end{equation}
for all $m\in{\mathcal N}$.\\

Secondly, we will apply the standard dilation procedure to the special operation $B$ considered in (\ref{KrausB})
in order to obtain a corresponding standard dilation for the stochastic matrix $T$. We obtain, for all $m,n\in{\mathcal N}$,
\begin{eqnarray}
\label{Tdila1}
  T_{mn} &\stackrel{(\ref{defTmn})}{=}& \langle m | \left( \sum_i A_i\,| n\rangle \langle n|\, A_i^\ast \right) |m\rangle\\
  \label{Tdila2}
   &\stackrel{(\ref{OI2})}{=}&\langle m |  A\left(|n\rangle \langle n| \right) |m\rangle\\
   &\stackrel{(\ref{Dila})}{=}&\langle m | \mbox{Tr}_2\left(U\left(  |n\rangle \langle n|\otimes |0\rangle \langle 0|\right) U^\ast\right) |m\rangle\\
   \label{Tdila1}\label{Tdila3}
   &=& \sum_i \langle m \,i  |  \left(U\left(  |n\rangle \langle n|\otimes |0\rangle \langle 0|\right) U^\ast\right)  |    m\, i\rangle\\
   \label{Tdila4}
   &=& \sum_i \langle m \,i  |   U  |n\,0\rangle \langle n\,0| U^\ast  |    m\, i\rangle
   =\sum_i \left|\langle m \,i  |   U  |n\,0\rangle \right|^2\\
   \label{Tdila5}
   &=& \sum_i R_{mi, n0}
   \;.
\end{eqnarray}
Here we have used the definition of the $N^2\times N^2$-matrix $R$ given by
\begin{equation}\label{defR}
 R_{mi,nj}:= \left|\langle m \,i  |   U  |n\,j\rangle \right|^2
 \;,
\end{equation}
for all $m,n,i,j\in {\mathcal N}$. $R$ is  bi-stochastic, and even \textit{uni-stochastic},
since its entries are the squares of the absolute values of the entries of a unitary matrix $U$.
Eq.~(\ref{Tdila5}) confirms that the standard dilation of the operation $B$ gives rise to
a special kind of an environment dilation of the stochastic matrix $T$ by means of a bi-stochastic matrix $R$,
where ${\mathcal M}={\mathcal N}$ and ${\boldsymbol\rho}=(1,0,\ldots,0)^\top$.
We correspondingly define:
\begin{defi}\label{D0}
 A bi-stochastic  $N^2\times N^2$-matrix $R$ is called a ``standard dilation" of a stochastic $N\times N$-matrix $T$ iff
 \begin{equation}\label{standil}
   T_{mn}=\sum_i R_{mi, n0}
   \,
 \end{equation}
 for all $m,n\in{\mathcal N}$ and some $0\in{\mathcal N}$.
 \end{defi}

 We have thus shown the following:
 \begin{theorem}\label{TSD}
   Every stochastic matrix $T$ admits a standard dilation by means of a bi-stochastic  $N^2\times N^2$-matrix $R$.
   Moreover, $R$ can be chosen as \textit{uni-stochastic}, i.~e.,
   such that the entries of $R$ can be written as the squares of the absolute values of the entries
   of some unitary matrix $U$.
 \end{theorem}
 It is worth noting that for dimensions greater than two not every bi-stochastic matrix will be uni-stochastic, see, e.~g., \cite{BE05}.

 Further, it is possible to directly describe the standard dilation of a stochastic matrix $T$, without recourse to the dilation of operations.
 To this end we define
 \begin{equation}\label{defRdir}
  R_{mi,nj}:=\left\{
  \begin{array}{r@{\quad : \quad}l}
  T_{mi}\,\delta_{in} & j=0,\\
   \frac{1}{N(N-1)}\,(1-T_{mi}) & \mbox{else},
  \end{array}
  \right.
 \end{equation}
 for all $m,i,n,j\in{\mathcal N}$.
 As for dilations of operations there is a considerable freedom in the choice of the dilation.
 In (\ref{defRdir}) we have made a particular simple choice resulting in a ``noisy" matrix $R$.
 It follows that
 \begin{equation}\label{standil1}
  \sum_i R_{mi, n0} \stackrel{(\ref{defRdir})}{=}   \sum_i T_{mi}\,\delta_{in}=T_{mn}
  \;,
 \end{equation}
 and hence (\ref{standil}) is satisfied. It remains to check that (\ref{defRdir}) defines a bi-stochastic $N^2\times N^2$-matrix $R$.
 Clearly, $R_{mi,nj}\ge 0$ for all $m,i,n,j\in{\mathcal N}$.  Moreover, for $j=0$,
 \begin{equation}\label{Rstochj0}
 \sum_{mi}R_{mi, nj} =  \sum_{mi}R_{mi, n0}\stackrel{(\ref{standil1})}{=}\sum_m T_{mn}=1
   \;,
 \end{equation}
 and, for $j\neq 0$,
 \begin{eqnarray}
 \label{Rstoch1}
    \sum_{mi}R_{mi, nj}&\stackrel{(\ref{defRdir}}{=}&\frac{1}{N(N-1)}\sum_{mi}\left( 1-T_{mi}\right)  \\
 \label{Rstoch2}
     &=& \frac{1}{N(N-1)} \left(N^2- \sum_i \underbrace{\left(\sum_m T_{mi} \right)}_{1} \right)=\frac{1}{N(N-1)} \left(N^2-N\right)=1
     \;.
 \end{eqnarray}
 Hence $R$ is left stochastic. To show that it is also right stochastic we consider
 \begin{eqnarray}
 \label{Rstoch3}
  \sum_{nj}R_{mi,nj} &=&  \sum_{n}R_{mi,n0}+ \sum_{n,j\neq 0}R_{mi,nj} \\
    &\stackrel{(\ref{defRdir})}{=}& \sum_n T_{mi}\,\delta_{in}+\frac{1}{N(N-1)}\sum_{n,j\neq 0}\left(1-T_{mi}\right)\\
    &=& T_{mi}+\left(1-T_{mi}\right)=1
    \;.
 \end{eqnarray}
 This completes the proof that (\ref{defRdir}) defines a standard dilation of the stochastic matrix $T$.

%%%%%%%%%%%%%%%%%%%%%%%%%%%%%%%%%%%%%%%%%%%%%%%%%%%%%%%%%%%%%%%%%%%%%%%%%%%%%%%%%%%%%%%%%%%%%%%%%%%%%%%%%%%%%%%%%%%%%%%%%%%%%%%%%%%%%%%%%%
\section{Comparison of two dilations of stochastic matrices}\label{sec:C2D}
%%%%%%%%%%%%%%%%%%%%%%%%%%%%%%%%%%%%%%%%%%%%%%%%%%%%%%%%%%%%%%%%%%%%%%%%%%%%%%%%%%%%%%%%%%%%%%%%%%%%%%%%%%%%%%%%%%%%%%%%%%%%%%%%%%%%%%%%%%

We have obtained two seemingly different forms of dilations (\ref{XSY}) and (\ref{standil}) of stochastic matrices
in the Sections \ref{sec:DCG}, Theorem \ref{T1}, and \ref{sec:DO1}, Theorem \ref{TSD}, and it remains to analyze how they are related.

First, we will extend the definition of a dilation of a stochastic matrix $T$ by means of uniform coarse graining and define:
\begin{defi}\label{D1}
Let $T$ be a stochastic $N\times N$-matrix.
 A  ``dilation of $T$ by coarse graining" is a quadruple $({\mathcal D},X,S,Y)$, where
 ${\mathcal D}$ is a finite set of size $d$, ${\mathcal D}= \biguplus_{n\in{\mathcal N}}{\mathcal D}_n$,
 $X:\Delta(d) \rightarrow \Delta(N)$ is defined by (\ref{matX}),
 $Y:\Delta(N) \rightarrow \Delta(d)$ is an \textit{arbitrary} affine right inverse of $X$,
 and $S:\Delta(d)\rightarrow\Delta(d)$ a bi-stochastic matrix such that
 \begin{equation}\label{defdil}
   T= X\,S\,Y
 \end{equation}
 holds.
\end{defi}

The ``dilation by uniform coarse graining" is a special case of Definition \ref{D1} since the uniform right inverse $Y$ defined
by (\ref{defY}) is an instance of a right inverse of $X$.
Then it is clear that, according to Theorem \ref{T1}, every stochastic matrix $T$ with a rational fixed point admits a dilation $({\mathcal D},X,S,Y)$
by coarse graining.
However, we will show that also Definition \ref{D3} can be viewed as a special case of Definition \ref{D1}.
To this end we assume that an environmental dilation $\left({\mathcal M},{\boldsymbol\rho},R\right)$ of $T$ is given
and define ${\mathcal D}:= {\mathcal N}\times {\mathcal M}$, hence $d=N\times M$. The partition
${\mathcal D}= \biguplus_{n\in{\mathcal N}}{\mathcal D}_n$ is then chosen as the projection onto the first factor, i.~e., by
${\mathcal D}_n:= \{n\} \times{\mathcal M}=\{(n,i)\left| i\in{\mathcal M}\right.\}$ for all $n\in{\mathcal N}$.
It follows that $X:\Delta(d) \rightarrow \Delta(N)$ is just the map of a probability distribution $\mathbf{P}$
onto its first marginal probability ${\sf M}_1\mathbf{P}$
defined by $\left({\sf M}_1\mathbf{P}\right)_n=\sum_i P_{n,i}$. In components: $X_{m,k i}=\delta_{mk}$ for all $i\in{\mathcal M}$ which implies
\begin{equation}\label{XP}
 \left( X\,\mathbf{P}\right)_n=\sum_{ki} X_{n,k i}\,P_{k,i}=\sum_{ki} \delta_{nk}\,P_{k,i}=\sum_i P_{n,i}=\left({\sf M}_1\mathbf{P}\right)_n
 \;.
\end{equation}
For the right inverse $Y$ of $X$ we do not choose an analogue of (\ref{defY}) but rather
$Y_{mi,k}:=\delta_{m,k}\,\rho_i$. This implies
\begin{equation}\label{Pp}
  P_{m,i}:= (Y \mathbf{p})_{mi}= \sum_{k}Y_{mi,k}\,p_k=\sum_{k}\delta_{m,k}\,\rho_i\,p_k=p_m\,\rho_i
  \;.
\end{equation}
Hence  $\mathbf{P}=Y \mathbf{p}$ is the product probability distribution  $\mathbf{p}\otimes{\boldsymbol\rho}$.
It is clear that
$X\,Y\, \mathbf{p}=X \mathbf{P}= {\sf M}_1\left( \mathbf{p}\otimes{\boldsymbol\rho} \right)= \mathbf{p}$
and hence $Y$ is a right inverse of $X$.

It remains to show $T= X\,R\,Y$ with $X$ and $Y$ as defined above and $R$ being the bi-stochastic matrix of the
environmental dilation:
\begin{equation}\label{XRY1}
 \left(X\,R\,Y\right)_{mn}=\sum_{k\ell r s}X_{m,k\ell}\,R_{k\ell,rs}\,Y_{rs,n}
 =\sum_{k\ell r s}\delta_{m,k}\,R_{k\ell,rs}\,\delta_{r,n}\,\rho_s
 =\sum_{\ell s} R_{m\ell,n s}\,\rho_s\stackrel{(\ref{D3coor})}{=}T_{mn}
 \;,
\end{equation}
for all $m,n\in{\mathcal N}$. Thus we have proven the following:
\begin{prop}\label{P1}
 Every environmental dilation of a stochastic matrix $T$ in the sense of Definitions \ref{D3}
 can be viewed as a dilation by coarse graining in the sense of Definition \ref{D1}.
\end{prop}

There is a fundamental difference between bi-stochastic matrices $R$ and proper stochastic matrices $T$:
Upon application of $S$ the Shannon entropy of a probability distribution is always non-decreasing, whereas $H(T{\mathbf p})$
may be smaller or larger than $H({\mathbf p})$, depending on ${\mathbf p}$.

The fact that the application of a bi-stochastic matrix $R$ does not decrease the entropy, see, e.~g., \cite[Prop.3.1]{PP11},
can be seen as follows:
Every bi-stochastic matrix $R$ can be written as a convex sum of permutational matrices.
This is the Birkhoff-von Neumann theorem, see \cite{B46,vN53}. The Shannon entropy is invariant under permutations,
but increases under a convex sum of probability distributions.  The latter is due to the concavity of the Shannon entropy, see, e.~g.,
\cite[ Ex.~$11.21$]{NC00}.

For proper stochastic matrices $T$ the entropy balance is different. This can be seen by means of the environmental dilation
$\left({\mathcal M},{\boldsymbol\rho},R\right)$ of $T$ that, according to Proposition \ref{P1}, is of the form $T=X\,R\,Y$.
Starting with a probability distribution ${\mathbf p}$ with entropy $H({\mathbf p})$ we first apply $Y$ with the result
$Y\,{\mathbf p}={\mathbf p}\otimes {\boldsymbol \rho}$ and a possibly increased entropy
$H({\mathbf p}\otimes {\boldsymbol\rho})=H({\mathbf p})+H({\boldsymbol \rho})\ge H({\mathbf p})$.
In the next application of the bi-stochastic $R$ the entropy is possibly further increased, $H({R\,Y\,\mathbf p})\ge H({\mathbf p})$.
In contrast, the final application of $X={\sf M}_1$ will decrease the entropy according to Proposition \ref{PS}. Hence the total
entropy balance of $T$ depends on the relation between the increase and decrease in entropy of the individual steps considered above.

\begin{figure}[t]
\centering
\includegraphics[width=0.7\linewidth]{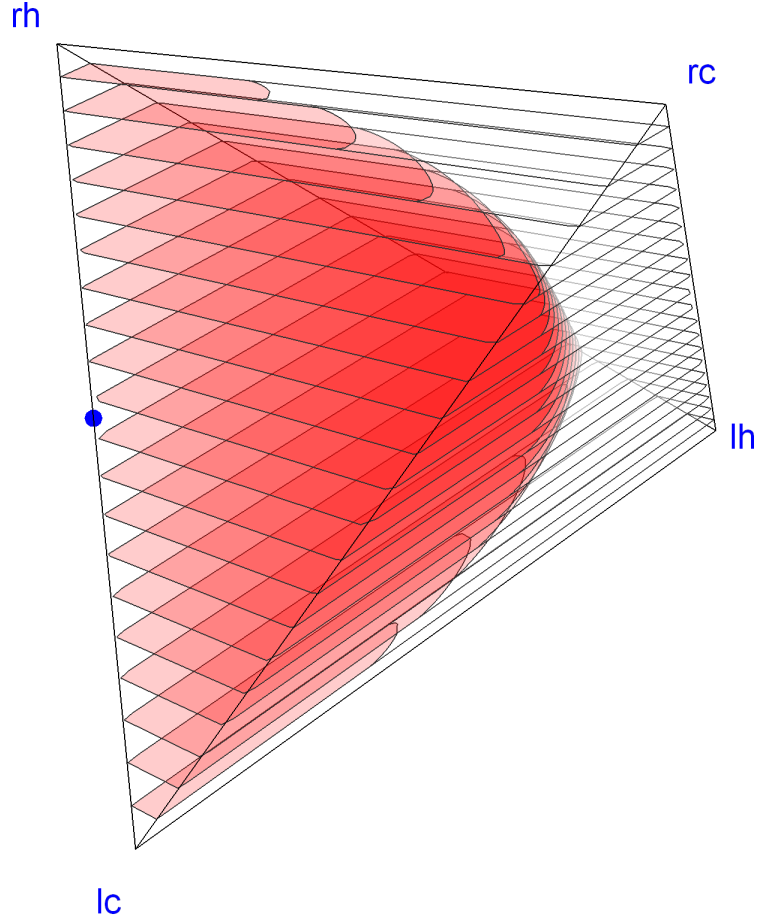}
\caption{Convex (red) subset $D(T)$ of the simplex $\Delta(4)$ corresponding to probability distributions
where application of the stochastic matrix $T$ defined by (\ref{MDT}),
describing the conditional action of Maxwell's demon, decreases the entropy.
The four vertices of $\Delta(4)$
are denoted by ``rh, rc, lh, lc" corresponding to the right/left and hot/cold states of a single molecule as explained in the text.
$D(T)$ is represented by its intersection with certain vertical planes in such a way that it becomes clear that its
boundary is composed of a smooth, strictly convex part and parts of the faces of $\Delta(4)$.
The asymptotic probability distribution $\mathbf{p}_\infty$ according to (\ref{limT})
is marked by a blue dot. It lies on the edge $\overline{rh,lc}$ representing all fixed points of $T$.
}
\label{FIGG4}
\end{figure}

Actually, the set $D(T)$ of probability distributions on which $T$ acts entropy decreasing, defined by
$D(T):=$
$\{{\mathbf p}\in \Delta(N) \left| H(T\,{\mathbf p})\le H({\mathbf p})\right.\}$,
is a non-empty convex subset of $\Delta(N)$, as follows from \cite[Prop.2]{S02}.
Its boundary is composed of a smooth, strictly convex surface given by the equation
$ H(T\,{\mathbf p})=H({\mathbf p})$ and possibly parts of the faces of $\Delta(N)$.
An example for $N=4$ is shown in Figure
\ref{FIGG4}. The barycenter $\frac{1}{N}{\mathbf e}$ of $\Delta(N)$ is always an element of $D(T)$ since it has maximal entropy.
The convex set  $F(T):=\{{\mathbf p}\in \Delta(N) \left| {\mathbf p}=T({\mathbf p})\right.\}$ of fixed points of $T$
is a face of $D(T)$. In the example shown in Figure \ref{FIGG4}, $F(T)$ is the edge spanned by the vertices ```rh" and ``lc".
For the bi-stochastic limit of $D(T)$ see \cite{PP11}.

%%%%%%%%%%%%%%%%%%%%%%%%%%%%%%%%%%%%%%%%%%%%%%%%%%%%%%%%%%%%%%%%%%%%%%%%%%%%%%%%%%%%%%%%%%%%%%%%%%%%%%%%%%%%%%%%%%%%%%%%%%%%%%%%%%%%%%%%%%
\section{Examples}\label{sec:EX}
%%%%%%%%%%%%%%%%%%%%%%%%%%%%%%%%%%%%%%%%%%%%%%%%%%%%%%%%%%%%%%%%%%%%%%%%%%%%%%%%%%%%%%%%%%%%%%%%%%%%%%%%%%%%%%%%%%%%%%%%%%%%%%%%%%%%%%%%%%

%%%%%%%%%%%%%%%%%%%%%%%%%%%%%%%%%%%%%%%%%%%%%%%%%%%%%%%%%%%%%%%%%%%%%%%%%%%%%%%%%%%%%%%%%%%%%%%%%%%%%%%%%%%%%%%%%%%%%%%%%%%%%%%%%%%%%%%%%%
\subsection{Dilation of a stochastic $2\times 2$-matrix}\label{sec:EX2}
%%%%%%%%%%%%%%%%%%%%%%%%%%%%%%%%%%%%%%%%%%%%%%%%%%%%%%%%%%%%%%%%%%%%%%%%%%%%%%%%%%%%%%%%%%%%%%%%%%%%%%%%%%%%%%%%%%%%%%%%%%%%%%%%%%%%%%%%%%
We consider the general stochastic $2\times 2$-matrix $T$ that is of the form
\begin{equation}\label{T2}
  T=\left(
\begin{array}{cc}
 1-b & a \\
 b & 1-a \\
\end{array}
\right)
\;,
\end{equation}
where $s$ and $b$ are parameters satisfying $0\le a,b\le 1$.
Its standard dilation in the sense of Definition \ref{D0} is given by the bi-stochastic $4\times 4$-matrix $R$:
\begin{equation}\label{R4}
 R=\left(
\begin{array}{cc|cc}
 1-b & 0 & \frac{b}{2} & \frac{b}{2} \\
 b & 0 & \frac{1-b}{2} & \frac{1-b}{2} \\
 \hline
 0 & a & \frac{1-a}{2} & \frac{1-a}{2} \\
 0 & 1-a & \frac{a}{2} & \frac{a}{2} \\
\end{array}
\right)
\;,
\end{equation}
where we have indicated the block structure of $R$ by lines.
One can directly check that $T$ is obtained from $R$ by adding the two left blocks of $R$.

For sake of comparison we also calculate the Sinkhorn representation \cite{S64} in the form $S= D^{(1)}\,T\,D^{(2)}$
such that $S$ is a bi-stochastic $2\times 2$-matrix. We assume $0<a,b<1$ in order to satisfy the conditions of Sinkhorn's theorem.
In this low-dimensional case the entries of the diagonal matrices $D^{(1)}$ and $D^{(1)}$
can be explicitly determined as:
\begin{eqnarray}
\label{D11}
  D^{(1)}_{11} &=& \sqrt{\frac{1-a}{1-b}}, \\
   D^{(1)}_{22} &=& \sqrt{\frac{a}{b}},\\
    D^{(2)}_{11} &=& \sqrt{\frac{1-a}{1-b}}\frac{1}{\sqrt{(1-a) (1-b)}+\sqrt{a b}}, \\
   D^{(2)}_{22} &=& \sqrt{\frac{(1-b) b}{(1-a) a}}\frac{1}{\sqrt{(1-a) (1-b)}+\sqrt{a b}}
   \;,
\end{eqnarray}
such that $S$  is given by
\begin{equation}\label{Smat}
  S=\left(
\begin{array}{cc}
 1-p & p \\
 p & 1-p \\
\end{array}
\right)
\;,
\end{equation}
where
\begin{equation}\label{pex}
  p=\frac{\sqrt{a b}}{\sqrt{(1-a) (1-b)}+\sqrt{a b}}
  \;,
\end{equation}
satisfying $0<p<1$.

%%%%%%%%%%%%%%%%%%%%%%%%%%%%%%%%%%%%%%%%%%%%%%%%%%%%%%%%%%%%%%%%%%%%%%%%%%%%%%%%%%%%%%%%%%%%%%%%%%%%%%%%%%%%%%%%%%%%%%%%%%%%%%%%%%%%%%%%%%
\subsection{Maxwell's demon}\label{sec:EXM}
%%%%%%%%%%%%%%%%%%%%%%%%%%%%%%%%%%%%%%%%%%%%%%%%%%%%%%%%%%%%%%%%%%%%%%%%%%%%%%%%%%%%%%%%%%%%%%%%%%%%%%%%%%%%%%%%%%%%%%%%%%%%%%%%%%%%%%%%%%

A stochastic matrix $T$ can be viewed as a description of a classical ``conditional action", see \cite{S20,S21}. Under certain
circumstances such a conditional action may decrease the Shannon entropy of the object system.
The environment dilation of $T$ can explain this apparent violation of the $2^{nd}$ law.
Coupling the object system to some auxiliary system (environment) and performing a bi-stochastic time evolution of the total system
does not decrease the total entropy but may lead to some entropy flow from the object system to the environment.

To illustrate this phenomenon we consider the thought experiment suggested by J.~C.~Maxwell, see, e.~g., \cite{EN98}, and
choose four states ${\mathcal N}=\{rh,rc,lh,lc\}$ of a single molecule that refer to its position
``right"$=r$ or ``left"$=l$ and its kinetic energy ``hot"$=h$ or ``cold"$=c$. The conditional action of Maxwell's demon is modelled by
a stochastic matrix of the form
\begin{equation}\label{MDT}
  T=\left(
\begin{array}{cccc}
 1 & 0 & \frac{1}{2} & 0 \\
 0 & \frac{1}{2} & 0 & 0 \\
 0 & 0 & \frac{1}{2} & 0 \\
 0 & \frac{1}{2} & 0 & 1 \\
\end{array}
\right)
\;.
\end{equation}
This means that the door between the ``right" and ``left" chamber
is kept closed and the molecule is left in its state if it is found to be ``right and hot" or ``left and cold".
However, if it is found in one of the states ``right and cold" or ``left and hot"
then the demon opens a small door and the molecule will change its position (but not its kinetic energy)
with probability $\frac{1}{2}$. The stochastic matrix (\ref{MDT}) has the set of fixed points
$F(T)=\{(\alpha,0,0,1-\alpha)^\top\left| 0\le \alpha \le 1\right.\}$.

If the initial probability distribution is uniform, $\mathbf{p}=\scriptsize{\frac{1}{4}}(1,1,1,1)^\top$,
with entropy $H(\mathbf{p}):=-\sum_n p_n \log p_n=\log 4=1.38629\ldots$, then after the conditional action the probability distribution will be
$\mathbf{q}=T\,\mathbf{p}=\scriptsize{\frac{1}{8}}(3,1,1,3)^\top$ with decreased entropy
$H(\mathbf{q})=\frac{3}{4} \log \left(\frac{8}{3}\right)+\frac{\log (8)}{4}\approx 1.25548$.
After applying $T$ repeatedly the initial probability distribution $\mathbf{p}$ will converge to the fixed point
\begin{equation}\label{limT}
  \lim_{n\to\infty} T^n \, \mathbf{p} ={\scriptsize{\frac{1}{2}}}(1,0,0,1)^\top=:\mathbf{p}_\infty
  \;,
\end{equation}
with an even lower entropy of $H(\mathbf{p}_\infty)=\log 2\approx 0.693147$.

We will apply to $T$ the standard dilation procedure defined in (\ref{defRdir}). This yields a bi-stochastic $16 \times 16$-matrix $R$
of the form
\begin{equation}\label{MaxR}
  R=\frac{1}{24}
  \left(
\begin{array}{cccccccccccccccc}
 24 & 0 & 0 & 0 & 0 & 0 & 0 & 0 & 0 & 0 & 0 & 0 & 0 & 0 & 0 & 0 \\
 0 & 0 & 0 & 0 & 2 & 2 & 2 & 2 & 2 & 2 & 2 & 2 & 2 & 2 & 2 & 2 \\
 0 & 0 & 0 & 0 & 2 & 2 & 2 & 2 & 2 & 2 & 2 & 2 & 2 & 2 & 2 & 2 \\
 0 & 0 & 0 & 0 & 2 & 2 & 2 & 2 & 2 & 2 & 2 & 2 & 2 & 2 & 2 & 2 \\
 0 & 0 & 0 & 0 & 2 & 2 & 2 & 2 & 2 & 2 & 2 & 2 & 2 & 2 & 2 & 2 \\
 0 & 12 & 0 & 0 & 1 & 1 & 1 & 1 & 1 & 1 & 1 & 1 & 1 & 1 & 1 & 1 \\
 0 & 0 & 0 & 0 & 2 & 2 & 2 & 2 & 2 & 2 & 2 & 2 & 2 & 2 & 2 & 2 \\
 0 & 12 & 0 & 0 & 1 & 1 & 1 & 1 & 1 & 1 & 1 & 1 & 1 & 1 & 1 & 1 \\
 0 & 0 & 12 & 0 & 1 & 1 & 1 & 1 & 1 & 1 & 1 & 1 & 1 & 1 & 1 & 1 \\
 0 & 0 & 0 & 0 & 2 & 2 & 2 & 2 & 2 & 2 & 2 & 2 & 2 & 2 & 2 & 2 \\
 0 & 0 & 12 & 0 & 1 & 1 & 1 & 1 & 1 & 1 & 1 & 1 & 1 & 1 & 1 & 1 \\
 0 & 0 & 0 & 0 & 2 & 2 & 2 & 2 & 2 & 2 & 2 & 2 & 2 & 2 & 2 & 2 \\
 0 & 0 & 0 & 0 & 2 & 2 & 2 & 2 & 2 & 2 & 2 & 2 & 2 & 2 & 2 & 2 \\
 0 & 0 & 0 & 0 & 2 & 2 & 2 & 2 & 2 & 2 & 2 & 2 & 2 & 2 & 2 & 2 \\
 0 & 0 & 0 & 0 & 2 & 2 & 2 & 2 & 2 & 2 & 2 & 2 & 2 & 2 & 2 & 2 \\
 0 & 0 & 0 & 24 & 0 & 0 & 0 & 0 & 0 & 0 & 0 & 0 & 0 & 0 & 0 & 0 \\
\end{array}
\right)
\;.
\end{equation}

The entropy balance of the dilation is as follows: The initial state of the total system
is $\mathbf{P}=\mathbf{p}\otimes (1,0,0,0)^\top$ with initial entropy $H(\mathbf{P})=H(\mathbf{p})+H(1,0,0,0)=H(\mathbf{p})=\log 4\approx 1.38629$.
After application of $R$ we obtain a distribution
$\mathbf{Q}:= R\,\mathbf{P}={\scriptsize\frac{1}{8}}(2,0,0,0,0,1,0,1,1,0,1,0,0,0,0,2)^\top$ with increased entropy
$H(\mathbf{Q})= \scriptsize{\frac{1}{2}}\log 32\approx 1.73287$. The two marginal distributions of $\mathbf{Q}$
are $\mathbf{Q}^{(1)}=\mathbf{q}$, since $R$ is a dilation of $T$, and  $\mathbf{Q}^{(2)}=\mathbf{p}$, due to the special ``noisy"
choice of $R$. Hence by passing to the marginal distributions the total entropy is further increased to
\begin{equation}\label{entropy}
H(\mathbf{Q}^{(1)})+H(\mathbf{Q}^{(2)})=\left(\frac{3}{4} \log \left(\frac{8}{3}\right)+\frac{\log (8)}{4}\right)+\log 4
\approx 1.25548 + 1.38629\approx 2.64178
\;.
\end{equation}
We see that the entropy decrease of the object system is overcompensated
by the entropy increase of the auxiliary system.

%%%%%%%%%%%%%%%%%%%%%%%%%%%%%%%%%%%%%%%%%%%%%%%%%%%%%%%%%%%%%%%%%%%%%%%%%%%%%%%%%%%%%%%%%%%%%%%%%%%%%%%%%%%%%%%%%%%%%%%%%%%%%%%%%%%%%%%%%%
\section{Summary}\label{sec:SU}
%%%%%%%%%%%%%%%%%%%%%%%%%%%%%%%%%%%%%%%%%%%%%%%%%%%%%%%%%%%%%%%%%%%%%%%%%%%%%%%%%%%%%%%%%%%%%%%%%%%%%%%%%%%%%%%%%%%%%%%%%%%%%%%%%%%%%%%%%%
We have obtained a dilation theorem for arbitrary stochastic matrices $T$ that represents $T$ in the form $T=X\,S\,Y$ such that
$S$ is bi-stochastic, $X$ is an affine map describing coarse graining and $Y$ a suitable right inverse of $X$.
Although this theorem bears some resemblance to Sinkhorn's theorem, written in the form $T=D_1^{-1}\,S\,D_2^{-1}$, there are important differences,
that have also been illustrated in the example of general stochastic $2\times 2$-matrices:
First, in general, the bi-stochastic matrix $S$ lives in a larger probability space whereas in Sinkhorn's theorem it will have the same size as $T$.
Second, the matrices $X,\,S,\,Y$ can be explicitly defined, at least in the standard dilation, but the Sinkhorn representation depends on, e.~g.,
application of the Sinkhorn-Knopp algorithm \cite{SK67}.
Moreover, Sinkhorn's theorem and its generalizations are restricted by certain positivity assumptions.

As an application of our result we have discussed the entropy balance for stochastic matrices $T$ and given an example where $T$ represents
a simplified form of the conditional action of Maxwell's demon. But the most important achievement of the concept of dilation will probably
lie in its analogy with the quantum mechanical notion of environment dilation of operations. Similar as in quantum mechanics, we can ``explain"
the occurrence of a particular form of state transformations by embedding the object system in a larger one, object system plus environment,
and by considering a restricted form of state evolution in the larger system. This restricted form will be given by a bi-stochastic matrix,
analogously to unitary time evolution in quantum mechanics. Note, that the strict analogy to unitary state evolution would consist of permutations;
by rather considering bi-stochastic matrices we have already admitted additional statistical mixtures via the Birkhoff-von Neumann theorem.
Finally, we constrain the obtained total state to the object system by forming the marginal distribution,
analogous to the partial trace in the quantum case.

We have also shown that the relation of our dilation concept to the analogous term in the quantum case is more than a mere analogy,
but follows as a special case, since all stochastic matrices can be obtained by
certain operations of the form $B= L\,A\,L$, see section \ref{sec:SMO}.
This fact makes our results relevant also for applications in quantum theory.

\section*{Acknowledgment}
%%%%%%%%%%%%%%%%%%%%%%%%%%%%%%%%%%%%%%%%%%%%%%%%%%%%%%%%%%%%%%%%%%%%%%%%%%%%%%%%%%%%%%%%%%%%%%%%%%%%%%%%%%%%%%%%%%%%%%%%%%%%%%%%%%%%%%%%%

I sincerely thank the members of the DFG Research Unit FOR2692 and Thomas Br\"ocker for fruitful discussions.

%\appendix

%%%%%%%%%%%%%%%%%%%%%%%%%%%%%%%%%%%%%%%%%%%%%%%%%%%%%%%%%%%%%%%%%%%%%%%%%%%%%%%%%%%%%%%%%%%%%%%%%%%%%%%%%%%%%%%%%%%%%%%%%%%%%%%%%%%%%%%%

\end{document}